\documentclass[aps,pre,showkeys,
reprint,                
twocolumn,              
nofootinbib,
floatfix]{revtex4-2}
\usepackage[OT4]{fontenc}
\usepackage[english]{babel}
\usepackage{tikz}
\usepackage{amsmath}
\usepackage{float}
\makeatletter
\let\newfloat\newfloat@ltx
\makeatother
\usepackage{algorithm}
\usepackage{algpseudocodex}
\usepackage{subcaption}
\usepackage{caption}
\usepackage{graphicx}
\usepackage{lipsum}
\usepackage{xcolor}
\usepackage{hyperref}
\hypersetup{
    colorlinks=true,
    urlcolor=blue,
}
\usepackage{cleveref}

\begin{document}

\author{Szymon Biernacki}
\author{Krzysztof Malarz}
\thanks{\includegraphics[width=10pt]{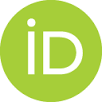}~\href{https://orcid.org/0000-0001-9980-0363}{0000-0001-9980-0363}}
\email{malarz@agh.edu.pl}
\affiliation{\mbox{AGH University of Science and Technology, Faculty of Physics and Applied Computer Science},\\
al. Mickiewicza 30, 30-059 Krak\'ow, Poland}

\title{Does social distancing matter for infectious disease propagation?
A SEIR model and Gompertz law based cellular automaton}

\begin{abstract}
In this paper, we present stochastic synchronous cellular automaton defined on a square lattice.
The automaton rules are based on the SEIR (susceptible $\to$ exposed $\to$ infected $\to$ recovered) model with probabilistic parameters gathered from real-world data on human mortality and the characteristics of the SARS-CoV-2 disease. 
With computer simulations, we show the influence of the radius of the neighborhood on the number of infected and deceased agents in the artificial population.
The increase in the radius of the neighborhood favors the spread of the epidemic.
However, for a large range of interactions of exposed agents (who neither have symptoms of the disease nor have been diagnosed by appropriate tests), even isolation of infected agents cannot prevent successful disease propagation.
This supports aggressive testing against disease as one of the useful strategies to prevent large peaks of infection in the spread of SARS-CoV-2-like disease. 
\end{abstract}

\date{\today}

\maketitle

\section{Introduction}

Currently, the death rate of SARS-CoV-2 \cite{Zhu_2001} in the whole world reached around 2\% of population \cite[tab `Closed Cases']{url:coronavirus}. Thus, one should not be surprised to the publication rash in this subject giving both: \emph{i}) theoretical bases of SARS-CoV-2 spreading, \emph{ii}) practical tips on preventing plague or even \emph{iii}) clinical case studies allowing easier to recognize and to treat cases of the disease.  The Web of Science database reveals over 80,000 and over 110,000 papers related to this topic registered in 2020 and in January-November 2021, respectively, in contrast to only 19 papers in 2019. Among them, only several \cite{ISI:000641801900016,ISI:000631029700056,ISI:000598519700004,ISI:000623111700017,ISI:000610760500001,ISI:000598669400008,ISI:000594821100019,ISI:000599877200011,ISI:000582762000094,ISI:000583404900051,ISI:000574548000001,ISI:000554890900001,ISI:000536109100001,ISI:000537095200003,Orzechowska_2018,PhysRevResearch.2.043379,Gabrick_2022} are based on cellular automata technique \cite{Ilachinski-2001,Wolfram-2002,Chopard-2005,Chopard-2012}.

The likely reason for this moderate interest in using this technique to simulate the spread of the COVID epidemic is the large degree of simplification of ``rules of the game'' in cellular automata. To fill this gap, in this work, we propose a cellular automaton based on a compartmental model, the parameters of which were adjusted to the realistic probabilities of the transitions between the states of the automaton. Let us note that modeling the spread of the epidemic is also possible with other models (see Refs. \onlinecite{deOliveira_2021,Lux_2021,Weisbuch_2021,Lorig_2021} for mini-reviews) including, for instance, those based on the percolation theory \cite{2101.00550}.

The history of the application of compartmental models to the mathematical modeling of infectious diseases dates to the first half of the 20-th century and works of Ross \cite{Ross_1916}, Ross and Hudson \cite{Ross_1917_1,Ross_1917_2}, Kermack and McKendrick \cite{Kermack1927,Kermack1991} and Kendall \cite{Kendall_1956}, see, for instance, Ref. \onlinecite{Hethcote_2000} for excellent review. 
In the compartmental model, the population is divided into several (usually labeled) compartments so that the agent only remains in one of them and the sequences of transitions between compartments (label changes) are defined. For instance, in the classical SIR model, agents change their states subsequently from {\em susceptible} ($\mathcal{S}$) via {\em infected} ($\mathcal{I}$) to {\em recovered} ($\mathcal{R}$) one \cite{Kermack1927,Kermack1991,PhysRevE.66.016128}.
Infected agents can transmit the disease to their susceptible neighbors ($\mathcal{S}\to\mathcal{I}$) with a given probability $p_1$. The infected agent may recover ($\mathcal{I}\to\mathcal{R}$) with probability $p_2$. After recovering, the agents are immune and they can no longer be infected with the disease. These rules may be described by a set of differential equations 
\begin{subequations}
\label{eq:SIR}
\begin{equation}
\dfrac{d n_{\mathcal S}}{dt} = -\langle k\rangle p_1 n_{\mathcal S} n_{\mathcal I},
\end{equation}
\begin{equation}
\dfrac{d n_{\mathcal I}}{dt} = \langle k\rangle p_1  n_{\mathcal S} n_{\mathcal I}-p_2 n_{\mathcal I},
\end{equation}
\begin{equation}
\dfrac{d n_{\mathcal R}}{dt} = p_2 n_{\mathcal I},
\end{equation}
\end{subequations}
where $\langle k\rangle$ is the mean number of agents' contacts in the neighborhood and $n_{\mathcal S}$, $n_{\mathcal I}$, $n_{\mathcal R}$ represent the fraction of susceptible, infected, recovered agents, respectively. Typically, the initial condition for \Cref{eq:SIR} is
\begin{subequations}
\label{eq:SIR0}
\begin{equation}
n_{\mathcal S}(t=0) = 1-n_0,
\end{equation}
\begin{equation}
n_{\mathcal I}(t=0) = n_0,
\end{equation}
\begin{equation}
n_{\mathcal R}(t=0) = 0,
\end{equation}
\end{subequations}
where $n_0$ is the initial fraction of infected agents (fraction of `Patients Zero').

The transition rates between states (i.e., probabilities $p_1$ and $p_2$) may be chosen arbitrarily or they may correspond to the reciprocal of agents' residence times in selected states. In the latter case, residence time may be estimated by clinical observations \cite{url:coronavirus-symptoms,url:coronavirus-incubation-period}.
The probability $p_1$ describes the speed of disease propagation (infecting rate) while the value of $p_2$ is responsible for the frequency of getting better (recovering rate).
In this approximation, the dynamics of the infectious class depends on the reproduction ratio:
\begin{equation}
\label{eq:R_0}
    R_0=\dfrac{\langle k\rangle p_1}{p_2}.
\end{equation}
The case of $R_0=1$ separates the phase when the disease dies out and the phase when the disease spreads among the members of the population.

Equations \eqref{eq:SIR} describe a mean-field evolution, which simulates a situation in which all agents interact directly with each other. In low-dimensional spatial networks, the mean-field dynamics \eqref{eq:SIR} is modified by
diffusive mechanisms \cite{Barthelemy_2011}. 
In a realistic situation, the diffusive mode of epidemic spreading is mixed with the mean-field dynamics, corresponding to nonlocal transmissions resulting from the mobility of agents \cite{Burda2020}.

We use the SEIR model \cite{Dureau2013,Faranda2020}, upon extending the SIR model, where an additional compartment (labeled $\mathcal{E}$) is available and it corresponds to agents in {\em exposed} state. The exposed agents are infected but unaware of it---they neither have symptoms of the disease nor have been diagnosed by appropriate tests.
This additional state requires splitting the transition rate $p_1$ into $p_{\mathcal E}$ and $p_{\mathcal I}$ corresponding to transition rates (probabilities) $\mathcal S\to\mathcal E$ after contact with the exposed agent in state $\mathcal E$ and $\mathcal S\to\mathcal E$ after contact with the infected agent in state $\mathcal I$, respectively.
We would like to emphasize that both exposed (in state $\mathcal E$) and infected (in state $\mathcal I$) agents {\em  may transmit} disease.

According to \Cref{eq:SIR}, after recovering, the convalescent in state $\mathcal R$ lives forever, which seems contradictory to the observations of the real world.
Although the Bible Book of Genesis (5:5-27; 9:29) mentions seven men who lived over 900 years, in modern society---thanks to public health systems (and sometimes in spite of them)---contemporary living lengths beyond one hundred years are rather rare.
Mortality tables \cite{USA_mortality_2003} show some correlations between probability of death and age \cite{Richmond_2021}.
This observation was first published by Gompertz in 1825 \cite{Gompertz_1825}.
According to Gompertz's law, mortality $f$ increases exponentially with age $a$ of the individual as
\begin{equation}
\label{eq:Gompertz}
f(a)\propto\exp(b(a+c)),
\end{equation}
where $b$ and $c$ are constants.
Moreover, as we mentioned in the first sentence of the Introduction, people can also die earlier than Gompertz's law implies.
For example, an epidemic of fatal diseases increases the mortality rate.
To take care of these factors in modeling disease propagation, we consider removing agents from the population.
This happens with agents' age-dependent probabilities $f_G$ and $f_C$ for healthy and ill people, respectively.
The removed individual is immediately replaced with a newly born baby.

In this paper, we propose a cellular automaton based simultaneously on SEIR model of disease propagation and Gompertz's law of mortality. 
In \Cref{sec:model}, the cellular automaton, its rules and the available site neighborhoods are presented.
\Cref{sec:results} is devoted to the presentation of the results of simulations based on the proposed cellular automaton.
Discussion of the obtained results (\Cref{sec:discussion}) and conclusions (\Cref{sec:conclusions}) end the paper.
We note Refs.~\onlinecite{Vandamme_2021,Rocha_Filho_2021}, where age-structured populations were also studied with SEIR-based and multicompartments models, respectively.

\section{\label{sec:model}Model}

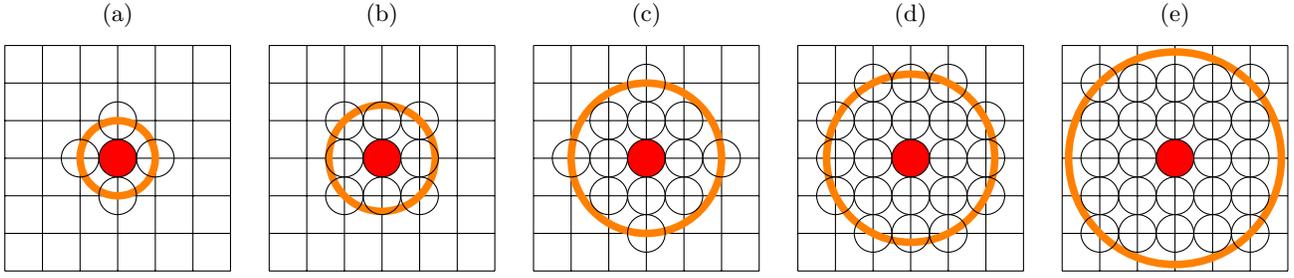
\begin{figure*}[tbp]
\centering
\begin{subfigure}[b]{.19\textwidth}
\caption{\label{fig:neighbourhood_r1.0}}
\begin{tikzpicture}[scale=0.5]
\draw (-3,-3) grid (3,3);
\draw[color=orange, line width=1mm] ( 0, 0) circle (1cm);
\draw[fill=red] ( 0, 0) circle (.5cm);
\draw (+1, 0) circle (.5cm);
\draw ( 0,+1) circle (.5cm);
\draw (-1, 0) circle (.5cm);
\draw ( 0,-1) circle (.5cm);
\end{tikzpicture}
\end{subfigure}
\begin{subfigure}[b]{.19\textwidth}
\caption{\label{fig:neighbourhood_r1.5}}
\begin{tikzpicture}[scale=0.5]
\draw (-3,-3) grid (3,3);
\draw[color=orange, line width=1mm] ( 0, 0) circle (1.4142cm);
\draw[fill=red] ( 0, 0) circle (.5cm);
\draw (+1, 0) circle (.5cm);
\draw ( 0,+1) circle (.5cm);
\draw (-1, 0) circle (.5cm);
\draw ( 0,-1) circle (.5cm);
\draw (+1,+1) circle (.5cm);
\draw (+1,-1) circle (.5cm);
\draw (-1,+1) circle (.5cm);
\draw (-1,-1) circle (.5cm);
\end{tikzpicture}
\end{subfigure}
\begin{subfigure}[b]{.19\textwidth}
\caption{\label{fig:neighbourhood_r2.0}}
\begin{tikzpicture}[scale=0.5]
\draw (-3,-3) grid (3,3);
\draw[color=orange, line width=1mm] ( 0, 0) circle (2cm);
\draw[fill=red] ( 0, 0) circle (.5cm);
\draw (+1, 0) circle (.5cm);
\draw ( 0,+1) circle (.5cm);
\draw (-1, 0) circle (.5cm);
\draw ( 0,-1) circle (.5cm);
\draw (+1,+1) circle (.5cm);
\draw (+1,-1) circle (.5cm);
\draw (-1,+1) circle (.5cm);
\draw (-1,-1) circle (.5cm);
\draw (+2, 0) circle (.5cm);
\draw ( 0,+2) circle (.5cm);
\draw (-2, 0) circle (.5cm);
\draw ( 0,-2) circle (.5cm);
\end{tikzpicture}
\end{subfigure}
\begin{subfigure}[b]{.19\textwidth}
\caption{\label{fig:neighbourhood_r2.5}}
\begin{tikzpicture}[scale=0.5]
\draw (-3,-3) grid (3,3);
\draw[color=orange, line width=1mm] ( 0, 0) circle (2.2360679cm);
\draw[fill=red] ( 0, 0) circle (.5cm);
\draw (+1, 0) circle (.5cm);
\draw ( 0,+1) circle (.5cm);
\draw (-1, 0) circle (.5cm);
\draw ( 0,-1) circle (.5cm);
\draw (+1,+1) circle (.5cm);
\draw (+1,-1) circle (.5cm);
\draw (-1,+1) circle (.5cm);
\draw (-1,-1) circle (.5cm);
\draw (+2, 0) circle (.5cm);
\draw ( 0,+2) circle (.5cm);
\draw (-2, 0) circle (.5cm);
\draw ( 0,-2) circle (.5cm);
\draw (+2,+1) circle (.5cm);
\draw (+1,+2) circle (.5cm);
\draw (-2,+1) circle (.5cm);
\draw (+1,-2) circle (.5cm);
\draw (+2,-1) circle (.5cm);
\draw (-1,+2) circle (.5cm);
\draw (-2,-1) circle (.5cm);
\draw (-1,-2) circle (.5cm);
\end{tikzpicture}
\end{subfigure}
\begin{subfigure}[b]{.19\textwidth}
\caption{\label{fig:neighbourhood_r3.0}}
\begin{tikzpicture}[scale=0.5]
\draw (-3,-3) grid (3,3);
\draw[color=orange, line width=1mm] ( 0, 0) circle (2.828427cm);
\draw[fill=red] ( 0, 0) circle (.5cm);
\draw (+1, 0) circle (.5cm);
\draw ( 0,+1) circle (.5cm);
\draw (-1, 0) circle (.5cm);
\draw ( 0,-1) circle (.5cm);
\draw (+1,+1) circle (.5cm);
\draw (+1,-1) circle (.5cm);
\draw (-1,+1) circle (.5cm);
\draw (-1,-1) circle (.5cm);
\draw (+2, 0) circle (.5cm);
\draw ( 0,+2) circle (.5cm);
\draw (-2, 0) circle (.5cm);
\draw ( 0,-2) circle (.5cm);
\draw (+2,+1) circle (.5cm);
\draw (+1,+2) circle (.5cm);
\draw (-2,+1) circle (.5cm);
\draw (+1,-2) circle (.5cm);
\draw (+2,-1) circle (.5cm);
\draw (-1,+2) circle (.5cm);
\draw (-2,-1) circle (.5cm);
\draw (-1,-2) circle (.5cm);
\draw (+2,+2) circle (.5cm);
\draw (-2,+2) circle (.5cm);
\draw (+2,-2) circle (.5cm);
\draw (-2,-2) circle (.5cm);
\end{tikzpicture}
\end{subfigure}
\caption{\label{fig:neighbourhoods} Sites in various neighborhoods $\mathcal N$ on a square lattice. (a) von Neumann's neighborhood ($r=1$, $z=4$), (b) Moore's neighborhood ($r=\sqrt{2}$, $z=8$), (c) neighborhood with sites up to the third coordination zone ($r=2$, $z=12$), (d) neighborhood with sites up to the fourth coordination zone ($r=\sqrt{5}$, $z=20$), (e) neighborhood with sites up to the fifth coordination zone ($r=2\sqrt{2}$, $z=24$).}
\end{figure*}

We use the cellular automata technique \cite{Ilachinski-2001,Wolfram-2002,Chopard-2005,Chopard-2012} to model disease propagation.
The cellular automata technique is based on several assumptions, including:
\begin{itemize}
    \item discrete (geometrical) space (i.e. regular lattice) and time;
    \item discrete and finite set of available states of the single lattice's site;
    \item local rule $\mathcal F$ of synchronous site states update.
\end{itemize}
The rule $\mathcal F$ defines the state $s_i$ of the site $i$ at time $(t+1)$ basing on this site state $s_i$ at time $t$ and the state of the sites in the $i$-th site's neighborhood $\mathcal N$
\[
s_i(t+1)=\mathcal F\Big( s_i(t);\mathcal N(s_i(t)) \Big).
\]

We adopt SEIR model for the simulation of disease spreading by probabilistic synchronous cellular automata on a square lattice with various neighborhoods $\mathcal N$. 
Every agent may be in one of four available states: susceptible ($\mathcal{S}$), exposed ($\mathcal{E}$), infected ($\mathcal{I}$) or recovered ($\mathcal{R}$).
The agents in $\mathcal{E}$ and $\mathcal{I}$ are characterized with different values of the probability of infection ($p_\mathcal{E}$ and $p_\mathcal{I}$) and different range of interactions (radii of neighborhoods $r_\mathcal{E}$ and $r_\mathcal{I}$).
The considered neighborhoods (and their radii) are presented in \Cref{fig:neighbourhoods}.

Initially (at $t=0$), every agent is in $\mathcal{S}$ state, their age $a$ is set randomly from normal distribution with mean value $\langle a\rangle=50\cdot 365$ days and dispersion $\sigma^2=25\cdot 365$ days. The `Patient Zero' in $\mathcal{E}$ state is placed randomly at a single site of a square lattice with $L^2=100^2$ nodes.

Every time step (which corresponds {\em literally} to a single day in real-world) the lattice is scanned in typewriter order to check the possible agent state evolution:
\begin{itemize}
\item The susceptible agent may be infected ($\mathcal{S}\to\mathcal{E}$) by each agent in state $\mathcal{E}$ or $\mathcal{I}$ present in his/her neighbourhood $\mathcal N$ with radius $r_\mathcal{E}$ or $r_\mathcal{I}\le r_\mathcal{E}$, respectively.
We set $r_\mathcal{I}\le r_\mathcal{E}$ as we assume that infected (and aware of the disease) agents are more responsible than exposed (unaware of the disease) ones.
The latter inequality comes from our assumption that the exposed agent is not careful enough in undertaking contacts with his/her neighbours while the infected agent is serious-minded and realizing the hazard of possible disease propagation and thus he/she avoids these contacts at least on the level assigned to exposed agents.  
The number and position of available neighbours who may infect the considered susceptible agent depend on the value of radius $r_\mathcal{E}$ and/or $r_\mathcal{I}$ as presented in \Cref{fig:neighbourhoods}. 
The infection of the susceptible agents occurs with probability $p_\mathcal{E}$ (after contacting with agent in state $\mathcal{E}$) or $p_\mathcal{I}$ (after contacting with agent in state $\mathcal{I}$), respectively.
\item The incubation (i.e., the appearance of disease symptoms) takes $\tau_\mathcal{E}$ days---every agent in $\mathcal{E}$ state is converted to infected state ($\mathcal{E}\to\mathcal{I}$) with probability $1/\tau_\mathcal{E}$.
The exposed agent may die with age-dependent probability $f_C(a)$.
In such a case, he/she is replaced ($\mathcal{E}\to\mathcal{S}$) with newly born agent ($a=0$).
\item The disease lasts for $\tau_\mathcal{I}$ days.
The ill agent (in state $\mathcal{I}$) may either die (and be replaced by a newly born child $\mathcal{I}\to\mathcal{S}$) with age specific probability $f_C(a)$ or he/she can recover (and gain resistance to disease $\mathcal{I}\to\mathcal{R}$) with probability $1/\tau_\mathcal{I}$.
\item A healthy agent ($\mathcal{S}$ or $\mathcal{R}$) may die with a chance given by age dependent probability $f_G(a)$.
In such a case, it is replaced with a newly born susceptible baby (in state $\mathcal{S}$ and in age of $a=0$).
\end{itemize}

\begin{algorithm}[htbp]
\caption{\label{alg:SEIR}Single time step in automaton}
\begin{algorithmic}[1]
\State $t\gets t+1$
\State tmp\_pop $\gets$ pop 
\ForAll{$i\in$ pop} 
    \State age[$i$] $\gets$ age[$i$]$+1$
    \If{tmp\_pop[$i$] = $\mathcal{S}$}
        \ForAll{$j\in$ pop, $j\ne i$}
            \If{tmp\_pop[$j$] = $\mathcal{I} \land \Vert i,j\Vert\leq r_\mathcal{I} \land$ random()$\leq p_\mathcal{I}$}
                \State pop[$i$]$\gets \mathcal{E}$ \Comment{$\mathcal{S}\to\mathcal{E}$ after contacting with $\mathcal{I}$}
                \State $n_\mathcal{S}\gets n_\mathcal{S}-1$
                \State $n_\mathcal{E}\gets n_\mathcal{E}+1$
                \State break
            \EndIf
            \If{tmp\_pop[$j$] = $\mathcal{E} \land \Vert i,j\Vert\leq r_\mathcal{E} \land$ random()$\leq p_\mathcal{E}$}
                \State pop[$i$]$\gets\mathcal{E}$ \Comment{$\mathcal{S}\to\mathcal{E}$ after contacting with $\mathcal{E}$}
                \State $n_\mathcal{S}\gets n_\mathcal{S}-1$
                \State $n_\mathcal{E}\gets n_\mathcal{E}+1$
                \State break
            \EndIf
        \EndFor
    \EndIf
    \If{tmp\_pop[$i$] = $\mathcal{E}\, \land$ random() $< 1/\tau_\mathcal{E}$} 
        \State pop[$i$]$\gets\mathcal{I}$ \Comment{$\mathcal{E}\to\mathcal{I}$}
        \State $n_\mathcal{E}\gets n_\mathcal{E}-1$
        \State $n_\mathcal{I}\gets n_\mathcal{I}+1$
    \EndIf
    \If{tmp\_pop[$i$] = $\mathcal{I}\, \land$ random() $< 1/\tau_\mathcal{I}$}
        \State pop[$i$]$\gets\mathcal{R}$ \Comment{Recovered}
        \State $n_\mathcal{I}\gets n_\mathcal{I}-1$
        \State $n_\mathcal{R}\gets n_\mathcal{R}+1$
    \EndIf
    \If{tmp\_pop[$i$] = $\mathcal{S}\, \lor$ tmp\_pop[$i$] = $\mathcal{R}$}
        \If{random() $< f_{G}($age[$i$])}
            \State pop[$i$]$\gets\mathcal{S}$ \Comment{Removed\ldots}
            \State age[$i$]$\gets 0$ \Comment{\ldots then Susceptible}
            \If{tmp\_pop[$i$]=$\mathcal{S}$} $n_\mathcal{S}\gets n_\mathcal{S}-1$ \EndIf
            \If{tmp\_pop[$i$]=$\mathcal{R}$} $n_\mathcal{R}\gets n_\mathcal{R}-1$ \EndIf
            \State $n_\mathcal{S}\gets n_\mathcal{S}+1$
        \EndIf
    \Else \Comment{tmp\_pop[$i$] = $\mathcal{E}\, \lor$ tmp\_pop[$i$] = $\mathcal{I}$}
        \If{random() $< f_{C}$(age[$i$])}
            \State pop[$i$]$\gets \mathcal{S}$ \Comment{Removed\ldots}
            \State age[$i$]$\gets 0$ \Comment{\ldots then Susceptible}
            \If{tmp\_pop[$i$]=$\mathcal{E}$} $n_\mathcal{E}\gets n_\mathcal{E}-1$ \EndIf
            \If{tmp\_pop[$i$]=$\mathcal{I}$} $n_\mathcal{I}\gets n_\mathcal{I}-1$ \EndIf
            \State $n_\mathcal{S}\gets n_\mathcal{S}+1$
            \State $n_\mathcal{D}\gets n_\mathcal{D}+1$ \Comment{cumulative number of deaths caused by infections}
        \EndIf
    \EndIf
    \State return$(t,n_\mathcal{S}/L^2,n_\mathcal{E}/L^2,n_\mathcal{I}/L^2,n_\mathcal{R}/L^2,n_\mathcal{D}/L^2)$
\EndFor
\end{algorithmic}
\end{algorithm}

The agents' state modifications are applied synchronously to all sites. 
A single time step (from $t$ to $t+1$) of the system evolution described above is presented in \Cref{alg:SEIR}.
The $L^2$-long vector variables \texttt{tmp\_pop[]} and \texttt{pop[]} represent the current population (at time $t$) and the population in the next time step ($t+1$), respectively.
The presence of two such variables in a model implementation is caused by the synchronicity of cellular automaton.
The $i$-th element of these vectors keeps information on the state (either $\mathcal{S}, \mathcal{E}, \mathcal{I}$ or $\mathcal{R}$) of the $i$-th agent in the population.
The age $a$ (measured in days) for the $i$-th agent is kept in $L^2$-long vector \texttt{age[]} and its $i$-th element is either incremented (line 4 of \Cref{alg:SEIR}) or reset (lines 28 and 35 of \Cref{alg:SEIR}) in the case of removal. 
The \texttt{random()} function returns a real pseudo-random number uniformly distributed in $[0,1)$.
The function $\Vert i,j\Vert$ measures the Euclidean distance between agents $i$ and $j$.
The age-dependent daily death probability functions $f_C(\cdot)$ and $f_G(\cdot)$---based on real-world data---are defined in the next paragraph in \Cref{eq:fG,eq:fC}.
The numbers $n_\mathcal{S}$, $n_\mathcal{E}$, $n_\mathcal{I}$ and $n_\mathcal{R}$ of agents in various states must be initialized at $t=0$ basing on initial conditions.
The cumulative number $n_\mathcal{D}$ of deceased agents---earlier either in the exposed ($\mathcal{E}$) or in the infected ($\mathcal{I}$) state---is incremented in line 39 of \Cref{alg:SEIR}.
The numbers $n_\mathcal{S}$, $n_\mathcal{E}$, $n_\mathcal{I}$, $n_\mathcal{R}$ and $n_\mathcal{D}$ are normalized to the total number $L^2$ of agents in the system before their return.

Basing on American data on annual death probability \cite{USA_mortality_2003} and assuming 365 days a year, we predict the daily death probability as
\begin{equation}
\label{eq:fG}
f_G(a)=184 \cdot 10^{-10}\exp(0.00023(a+40259)),
\end{equation}
where $a$ is the agent's age expressed in days.
The data follow Gompertz's exponential law of mortality \cite{Gompertz_1825,Makowiec_2001}.

\begin{figure}
    \centering
    \includegraphics[width=0.99\columnwidth]{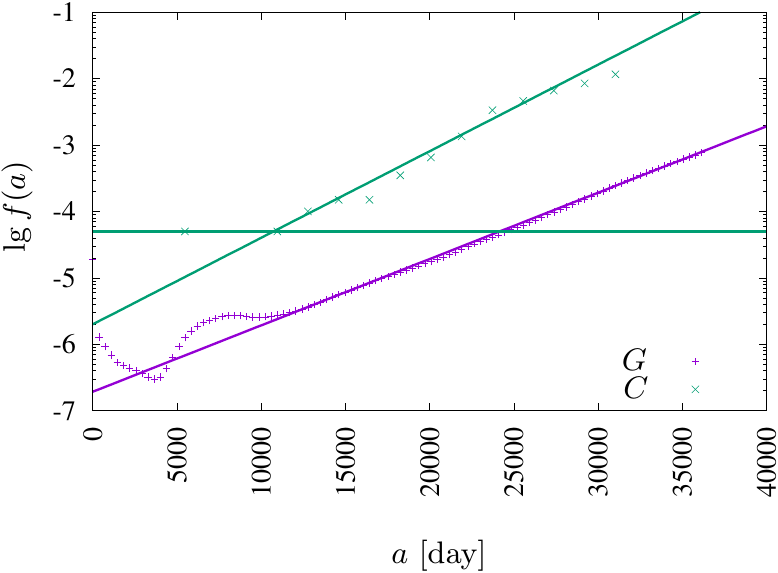}
    \caption{\label{fig:f}Daily death probability $f(a)$ for patients infected by the coronavirus (SARS-CoV-2, $\times$, $f_C(a)$, \cite{url:poland-by-age}) and natural death probability ($+$, $f_G(a)$, \cite{USA_mortality_2003}).}
\end{figure}

Using the same trick, we estimate the probability of daily death for infected people of age $a$ (expressed in days) as
\begin{equation}
\label{eq:fC}
    f_C(a)=
    \begin{cases}
    5 \cdot 10^{-5}                 & \iff a\le 30~y,\\
    2 \cdot 10^{-6} \exp(0.0003a)   & \iff a> 30~y.
    \end{cases}
\end{equation}
These probabilities are calculated as the chance of death during SARS-CoV-2 infection (based on Polish statistics \cite{url:poland-by-age}) divided by $(\tau_\mathcal{E}+\tau_\mathcal{I})$.
We assume that infection lasts $\tau_\mathcal{I}\approx 14$ days and that an incubation process takes $\tau_\mathcal{E}\approx 5$ days \cite{url:coronavirus-incubation-period}.
Exponential fits [\Cref{eq:fG,eq:fC}] to real-world data are presented in \Cref{fig:f}. 

\section{\label{sec:results}Results}

\begin{figure}[!htbp]
\centering
\begin{subfigure}[b]{.23\textwidth}
\caption{\label{fig:snap_r1.0}}
\includegraphics[width=.99\columnwidth]{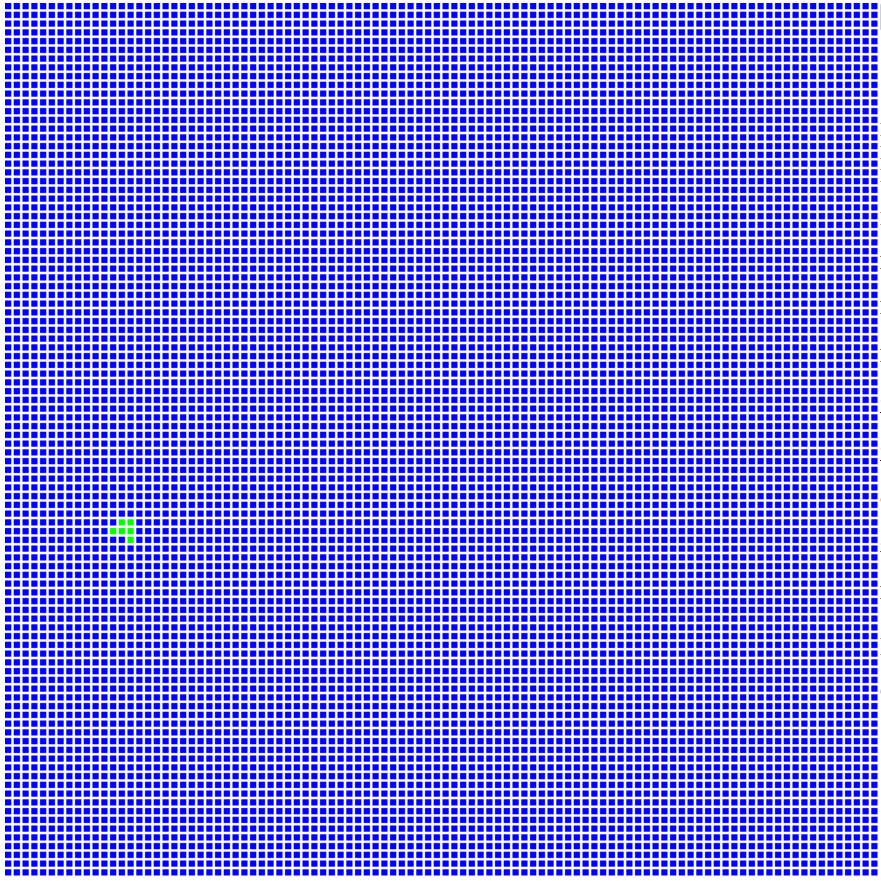}
\end{subfigure}
\begin{subfigure}[b]{.23\textwidth}
\caption{\label{fig:snap_r1.5}}
\includegraphics[width=.99\columnwidth]{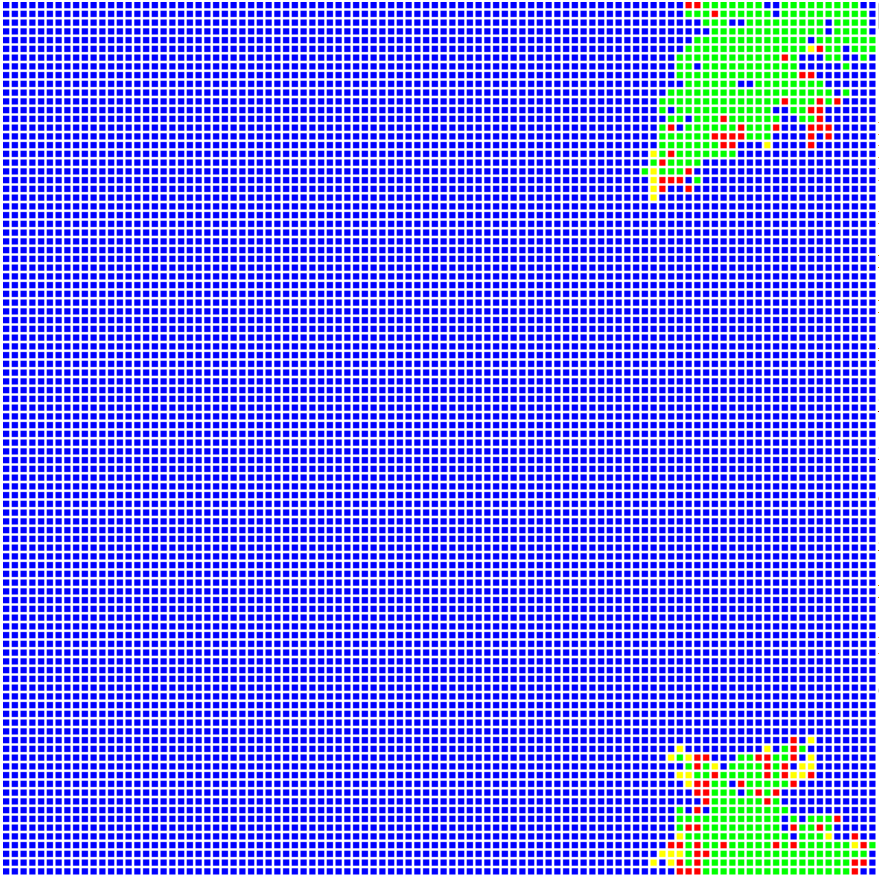}
\end{subfigure}
\begin{subfigure}[b]{.23\textwidth}
\caption{\label{fig:snap_r2.0}}
\includegraphics[width=.99\columnwidth]{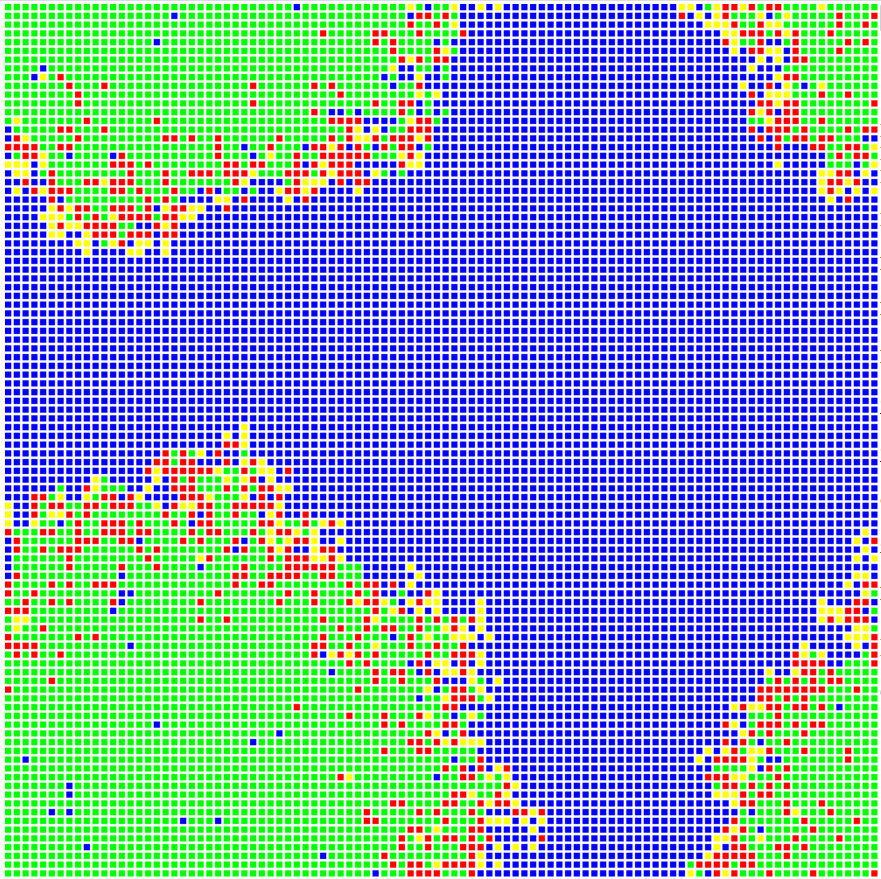}
\end{subfigure}
\begin{subfigure}[b]{.23\textwidth}
\caption{\label{fig:snap_r2.5}}
\includegraphics[width=.99\columnwidth]{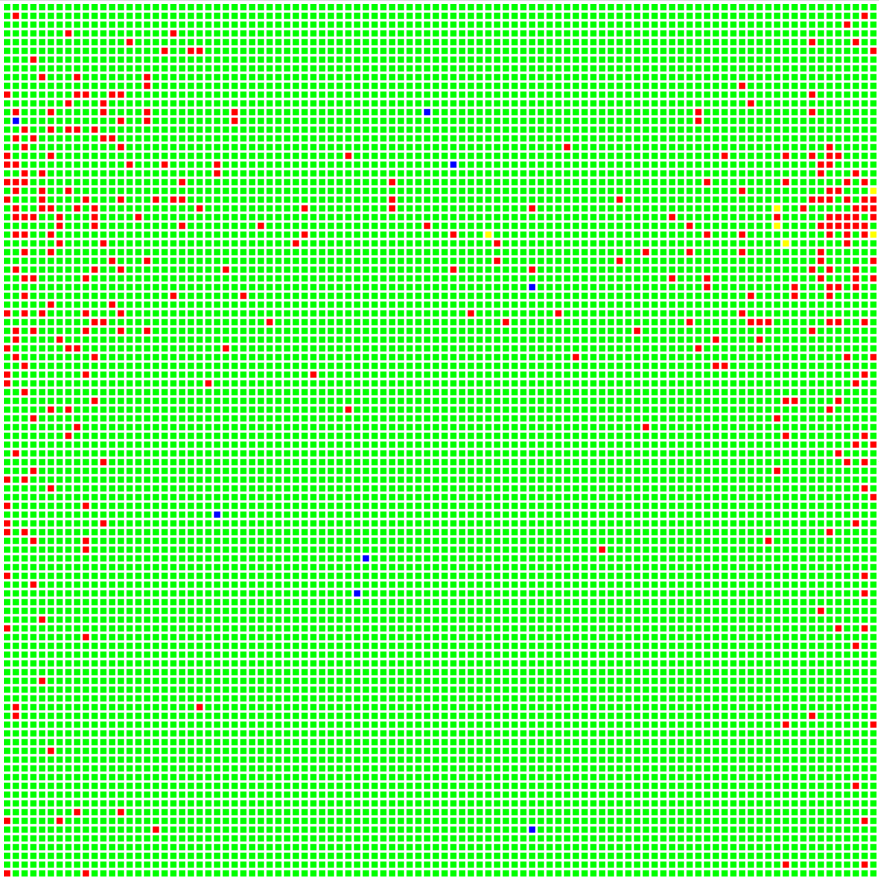}
\end{subfigure}
\begin{subfigure}[b]{.23\textwidth}
\caption{\label{fig:snap_r3.0}}
\includegraphics[width=.99\columnwidth]{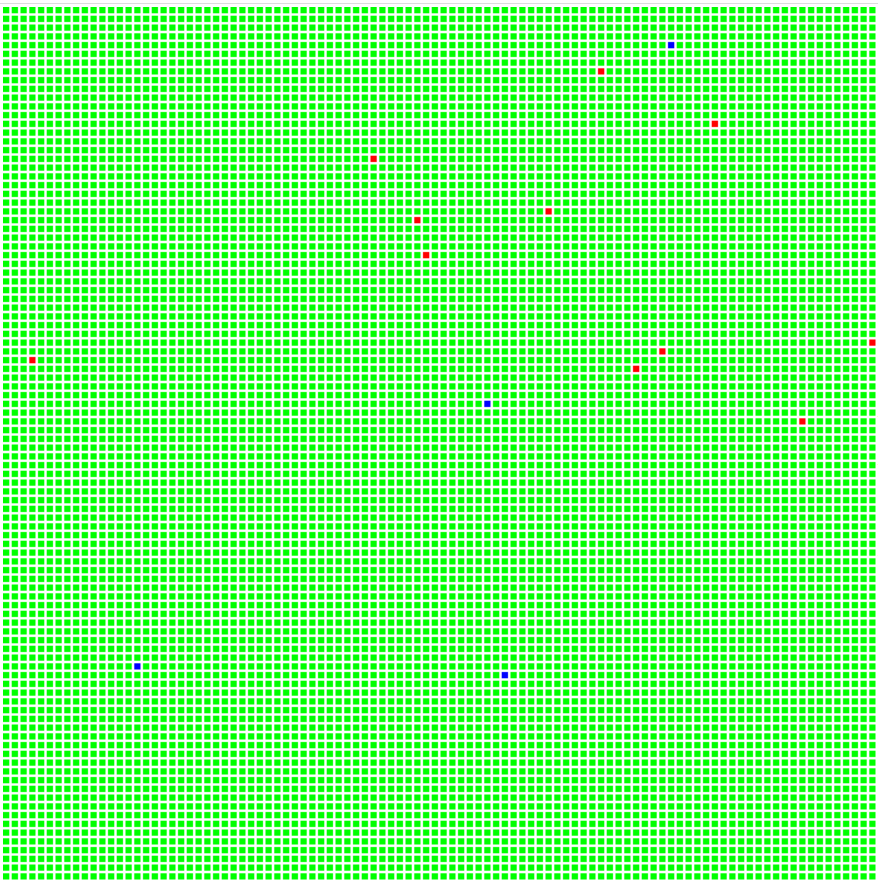}
\end{subfigure}
\begin{subfigure}[b]{.23\textwidth}
\caption{\label{fig:snap_long}}
\includegraphics[width=.99\columnwidth]{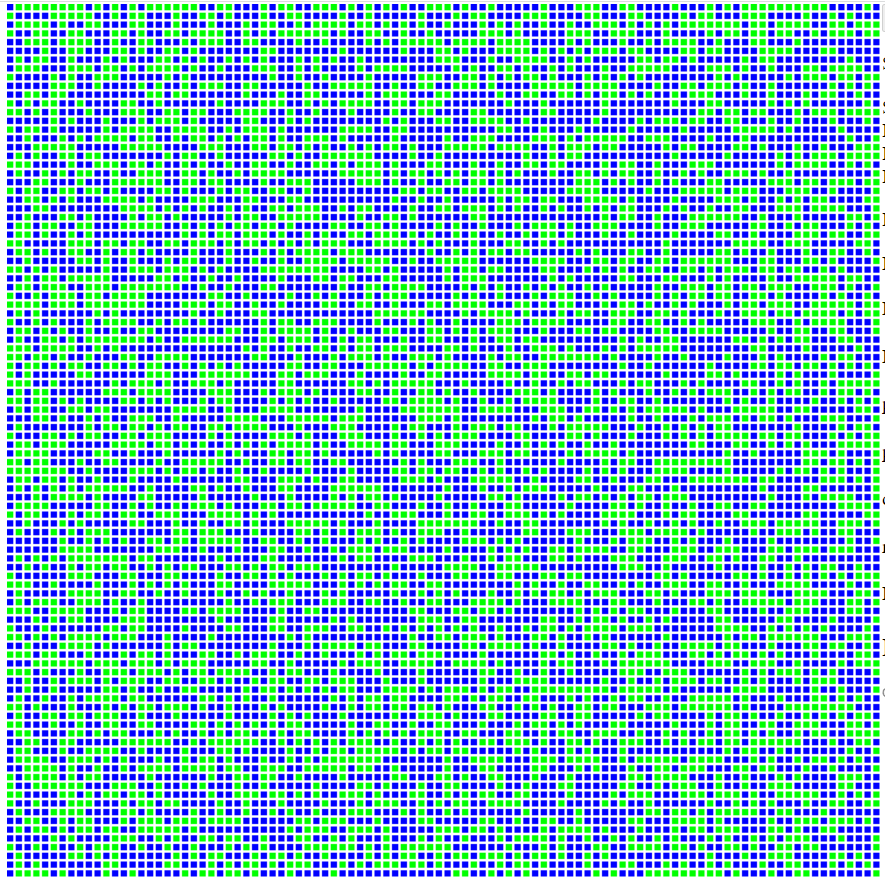}
\end{subfigure}
\caption{\label{fig:snapshots}Snapshots from direct simulation \cite{aplikacja} for  
$p_\mathcal{E}=0.03$, $p_\mathcal{I}=0.02$, $R=1$.
The assumed ranges of interactions are 
(a) $r_\mathcal{E}=r_\mathcal{I}=1$, 
(b) $r_\mathcal{E}=r_\mathcal{I}=1.5$,
(c) $r_\mathcal{E}=r_\mathcal{I}=2$,
(d, f) $r_\mathcal{E}=r_\mathcal{I}=2.5$,
(e) $r_\mathcal{E}=r_\mathcal{I}=3$.
The simulation takes $t=150$ time steps except of \Cref{fig:snap_long}, where situation after $t>20000$ time steps is presented.}
\end{figure}

In \Cref{fig:snapshots} snapshots from a single-run simulation are presented. They give a quantitative picture of the influence of the interaction range (neighborhood radius) on the spread of the disease. The snapshots in \Cref{fig:snap_r1.0,fig:snap_r1.5,fig:snap_r2.0,fig:snap_r2.5,fig:snap_r3.0} show the situation for fixed parameters $p_\mathcal{E}=0.03$ and $p_\mathcal{I}=0.02$ at the $t=150$ time step which corresponds to five months after introducing (at random site) `Patient Zero'. The last subfigure (\Cref{fig:snap_long}) presents a situation after a very long time of simulations ($t>20000$) where the recovered agents die due to their age (according to \Cref{eq:fG}) and are subsequently replaced by newly born children. 
For the interaction limited to the first coordination zone (\Cref{fig:neighbourhood_r1.0}) the disease propagation stays limited to the nearest-neighbors of the `Patient Zero'.
On the other hand, for the neighborhood with sites up to the fifth coordination zone (\Cref{fig:neighbourhood_r3.0}) for the same infection rates ($p_\mathcal{E}=0.03$ and $p_\mathcal{I}=0.02$) the disease affects all agents in the population (see \Cref{fig:snap_r3.0}).
The direct evolution of the system based on Ref.~\onlinecite{SBMSCThesis} can be simulated and observed with the JavaScript application available at \cite{aplikacja}.

\begin{figure}[htb]
\centering
\captionsetup{justification=centering}
\includegraphics[width=.99\columnwidth]{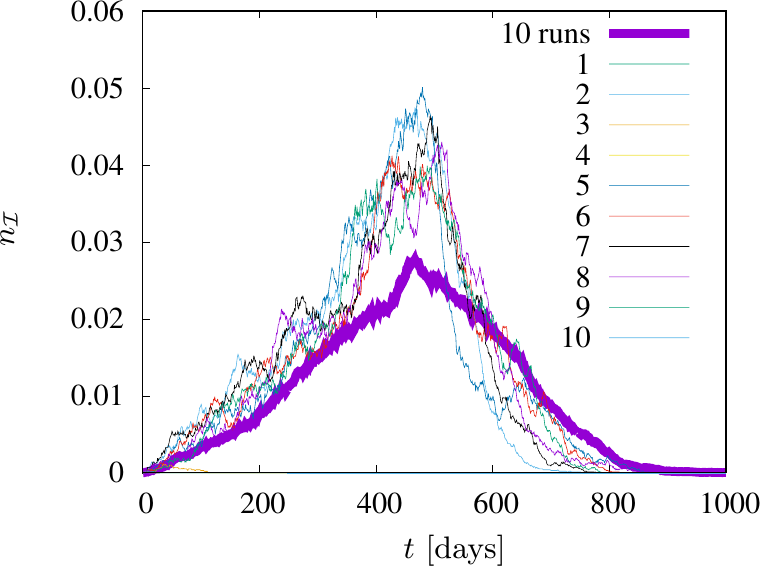}
\caption{\label{fig:I10Sim}Ten different simulations for values of neighbourhood radius $r_\mathcal{E}=r_\mathcal{I}$=1.5. $p_\mathcal{E} = 0.03$, $p_\mathcal{I} = 0.02$.}
\end{figure}

In \Cref{fig:I10Sim} the fraction $n_{\mathcal I}$ of infected agents (in state $\mathcal I$) is presented.
The figure shows the results of ten different simulations for values of neighborhood radius $r_\mathcal{E}=r_\mathcal{I}=1.5$, $p_\mathcal{E} = 0.03$, $p_\mathcal{I} = 0.02$.
In addition, the results of averaging over $R=10$ simulations are presented. 
In two out of ten cases, the epidemic died out right after the start, while in the remaining eight cases it lasted from about eight hundred to over a thousand time steps (days).
The figure also shows that the averaging of the results allows for a significant smoothing of the curves, which fluctuate strongly for individual simulations. 
Basing on this test [for (not shown) roughly twice large statistics, $R=25$, which do not reveal significant deviations], we decided to average our results (presented in \Cref{fig:pE0.005_pI0.005_rErI,fig:pE0.03_pI0.02,fig:maxIvsr}) over ten independent simulations.

The diagrams in \Cref{fig:pE0.005_pI0.005_rErI,fig:pE0.03_pI0.02} show the evolution of the epidemic.
Namely, they show the number of agents in each state on each day of the epidemic, as well as the cumulative fraction $n_{\mathcal D}$ of deaths ($\mathcal D$). 
The fraction $n_{\mathcal S}$ of susceptible agents and the fraction $n_\mathcal{R}$ of recovered agents are shown on the left vertical axis, while the fractions $n_{\mathcal E}$ of exposed agents and $n_{\mathcal I}$ of infected agents and the cumulative fraction $n_\mathcal D$ of deaths caused by infection are shown on the right vertical axis.

\subsection{\label{sec:rE_eq_rI_eq_0}$r_\mathcal{E}=r_\mathcal{I}=0$}

The case of $r_E=r_I=0$ (corresponding to total lockdown) leads to immediate disease dieout as only $n_0L^2$ `Patients Zero' at $t=0$ are infected and recover after about $\sim 1/q_C(a)$ time steps (days) depending on the agent's age $a$.

\subsection{\label{sec:rE_eq_rI}$r_\mathcal{E}=r_\mathcal{I}>0$}

\begin{figure}[!htbp]
\centering
\begin{subfigure}[b]{.63\columnwidth}
\caption{\label{fig:pE0.005_pI0.005_rErI1.0}$r_I=r_E=1$}
\includegraphics[width=0.94\columnwidth]{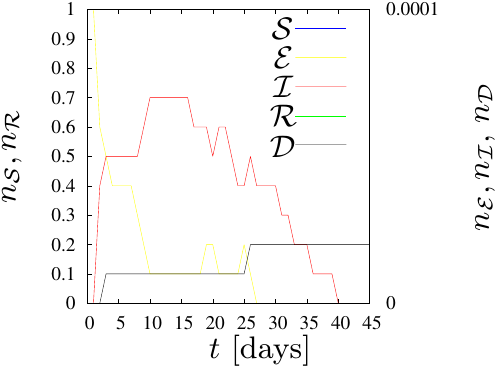}
\end{subfigure}
\begin{subfigure}[b]{.63\columnwidth}
\caption{\label{fig:pE0.005_pI0.005_rErI1.5}$r_I=r_E=1.5$}
\includegraphics[width=0.94\columnwidth]{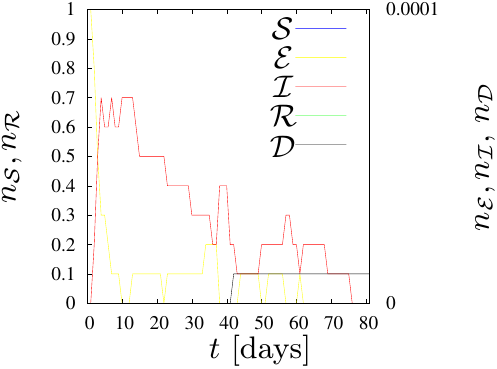}
\end{subfigure}
\begin{subfigure}[b]{.63\columnwidth}
\caption{\label{fig:pE0.005_pI0.005_rErI2.0}$r_I=r_E=2$}
\includegraphics[width=0.94\columnwidth]{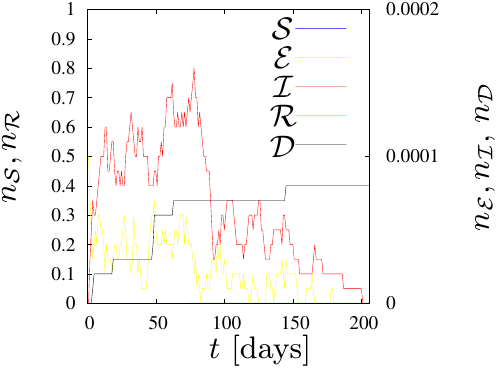}
\end{subfigure}
\begin{subfigure}[b]{.63\columnwidth}
\caption{\label{fig:pE0.005_pI0.005_rErI2.5}$r_I=r_E=2.5$}
\includegraphics[width=0.94\columnwidth]{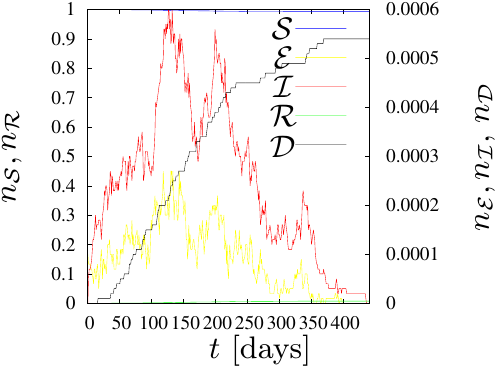}
\end{subfigure}
\begin{subfigure}[b]{.63\columnwidth}
\caption{\label{fig:pE0.005_pI0.005_rErI3.0}$r_I=r_E=3$}
\includegraphics[width=0.94\columnwidth]{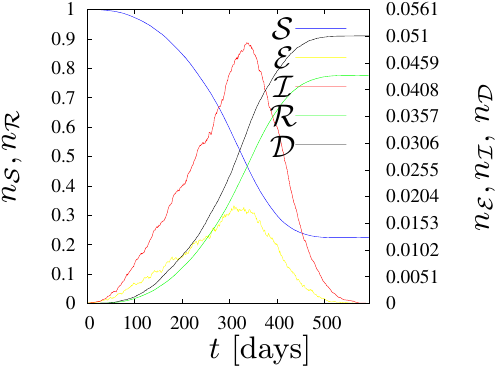}
\end{subfigure}
\caption{\label{fig:pE0.005_pI0.005_rErI}Dynamics of states fractions for various values of the neighborhood radius $r_\mathcal{E}=r_\mathcal{I}$. $p_\mathcal{E}=p_\mathcal{I}=0.005$, $R=10$.}
\end{figure}

In \Cref{fig:pE0.005_pI0.005_rErI} the dynamics of states fractions $n$ 
for various values of the neighborhood radius $r_{\mathcal E}=r_{\mathcal I}$ are presented. 
We assume infection rates $p_{\mathcal E}=p_{\mathcal I}=0.005$. 
The assumed transition rates $p_{\mathcal E}$ and $p_{\mathcal I}$ are very low.
As a result, the disease has a very limited chance of spreading in society.

\Cref{fig:pE0.005_pI0.005_rErI1.0} illustrates the situation where an infected person (independently either in the $\mathcal E$ or $\mathcal I$ states) can only infect the four closest neighbors. The epidemic lasts a maximum of forty days, the number of agents who are ill at the same time is less than one (on average, in ten simulations), there were only two deaths out of ten simulations, and the population was not affected by the disease.

The case presented in \Cref{fig:pE0.005_pI0.005_rErI1.5}, illustrates the situation in which each person can infect up-to eight neighbors. This does not cause significant changes during the epidemic compared with \Cref{fig:pE0.005_pI0.005_rErI1.0}, the average number of simultaneously ill agents remains below one, this time only one person in ten simulations died due to SARS-CoV-2 infection, and the duration of the epidemic was approximately 75 days---twice longer than presented in  \Cref{fig:pE0.005_pI0.005_rErI1.0}. We would like to emphasize that the term ``duration of the epidemic'' determines the time of the longest duration of the epidemic among the ten simulations carried out.

\Cref{fig:pE0.005_pI0.005_rErI2.0} shows the situation where there are twelve agents in the neighborhood of each cell. 
The transition rates ($p_{\mathcal E}$ and $p_{\mathcal I}$) turn out to be so low, that---despite extending the neighborhood---the epidemic vanishes quickly. 
This time the fractions of exposed and infected agents are slightly higher, the maximum number of sick agents in one day is more than one, the longest simulation lasted 80 days, and four agents died within ten iterations.

Increasing the radius of the neighborhood to 2.5 (see \Cref{fig:pE0.005_pI0.005_rErI2.5}) increased the number of exposed and infected agents more than twice compared to \Cref{fig:pE0.005_pI0.005_rErI2.0}. On the day of the peak of the epidemic, five agents were sick and seven died during the epidemic (on average). The epidemic lasted about 600 days, but only about 1\%--2\% of the population became infected throughout the epidemic.

Only an increase in the number of agents in the neighborhood to 24 (as presented in \Cref{fig:pE0.005_pI0.005_rErI3.0}) caused a smooth and rapid development of the epidemic. In this case, it lasted about 550 days, the largest number of sick agents in one day was approximately 500, and the same number of agents also died throughout the epidemic. During the epidemic, around 75\% of the population became infected.

\begin{figure*}[!htb]
\centering
\begin{subfigure}[b]{.329\textwidth}
\caption{\label{fig:pE0.03_pI0.02_rErI1.0}$r_\mathcal{I}=r_\mathcal{E}=1$}
\includegraphics[width=.94\columnwidth]{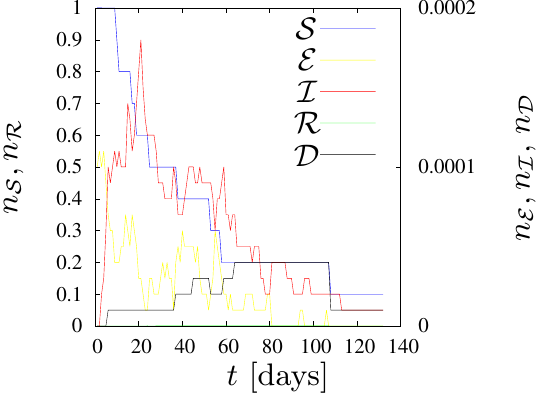}
\end{subfigure}
\begin{subfigure}[b]{.329\textwidth}
\caption{\label{fig:pE0.03_pI0.02_rI1.0_rE1.0}$r_\mathcal{E}=1$}
\includegraphics[width=.94\columnwidth]{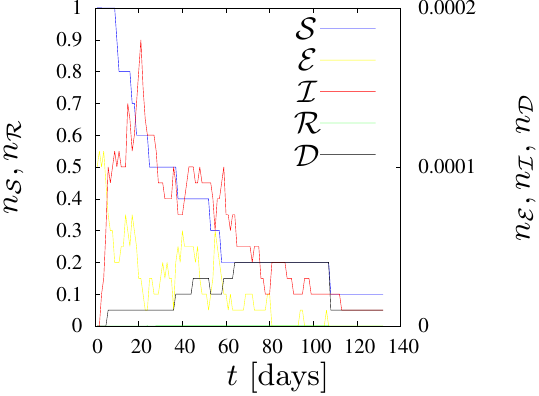}
\end{subfigure}
\begin{subfigure}[b]{.329\textwidth}
\caption{\label{fig:pE0.03_pI0.02_rE3.0_rI1.0}$r_\mathcal{I}=1$}
\includegraphics[width=.94\columnwidth]{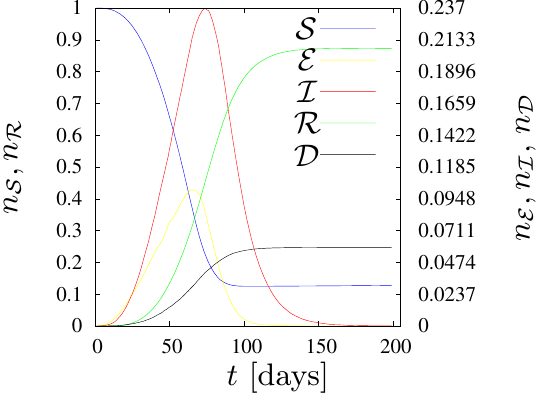}
\end{subfigure}
\begin{subfigure}[b]{.329\textwidth}
\caption{\label{fig:pE0.03_pI0.02_rErI1.5}$r_\mathcal{I}=r_\mathcal{E}=1.5$}
\includegraphics[width=.94\columnwidth]{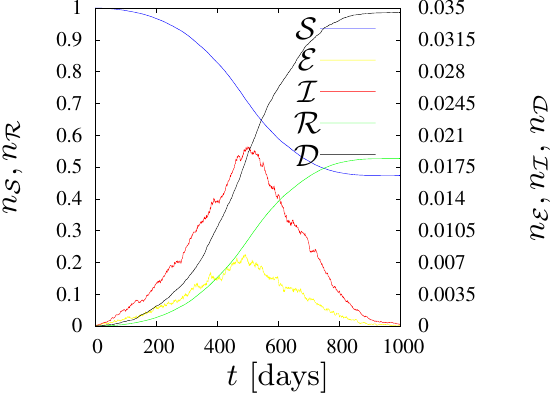}
\end{subfigure}
\begin{subfigure}[b]{.329\textwidth}
\caption{\label{fig:pE0.03_pI0.02_rI1.0_rE1.5}$r_\mathcal{E}=1.5$}
\includegraphics[width=.94\columnwidth]{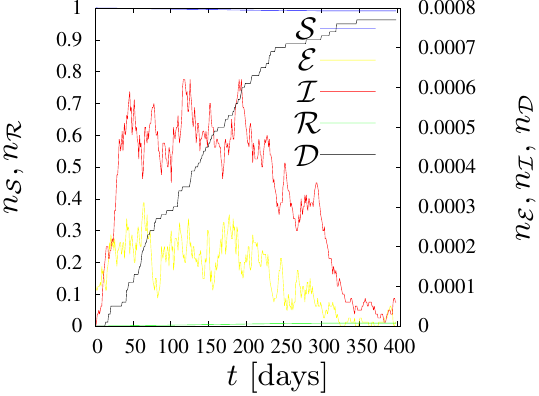}
\end{subfigure}
\begin{subfigure}[b]{.329\textwidth}
\caption{\label{fig:pE0.03_pI0.02_rE3.0_rI1.5}$r_\mathcal{I}=1.5$}
\includegraphics[width=.94\columnwidth]{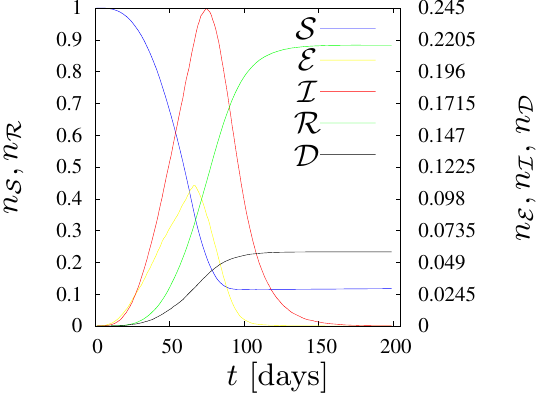}
\end{subfigure}
\begin{subfigure}[b]{.329\textwidth}
\caption{\label{fig:pE0.03_pI0.02_rErI2.0}$r_\mathcal{I}=r_\mathcal{E}=2$}
\includegraphics[width=.94\columnwidth]{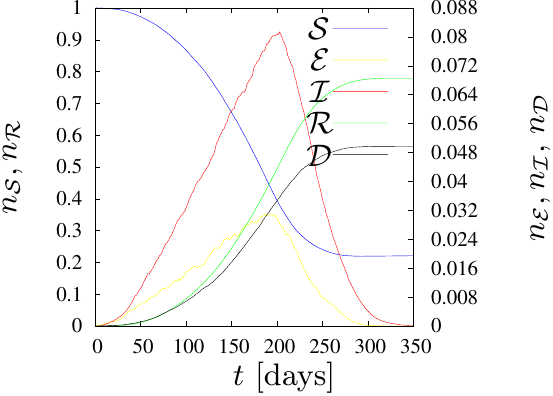}
\end{subfigure}
\begin{subfigure}[b]{.329\textwidth}
\caption{\label{fig:pE0.03_pI0.02_rI1.0_rE2.0}$r_\mathcal{E}=2$}
\includegraphics[width=.94\columnwidth]{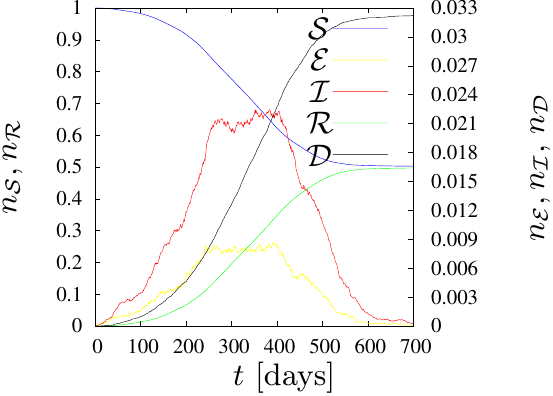}
\end{subfigure}
\begin{subfigure}[b]{.329\textwidth}
\caption{\label{fig:pE0.03_pI0.02_rE3.0_rI2.0}$r_\mathcal{I}=2$}
\includegraphics[width=.94\columnwidth]{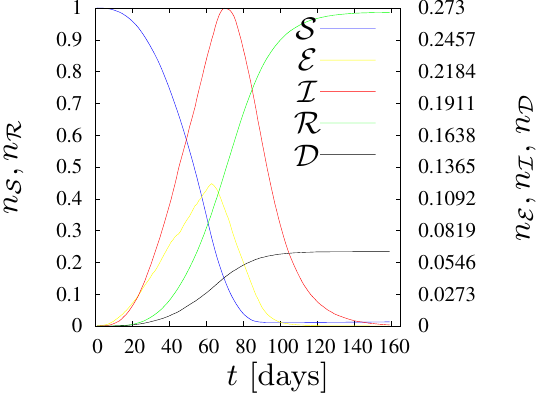}
\end{subfigure}
\begin{subfigure}[b]{.329\textwidth}
\caption{\label{fig:pE0.03_pI0.02_rErI2.5}$r_\mathcal{I}=r_\mathcal{E}=2.5$}
\includegraphics[width=.94\columnwidth]{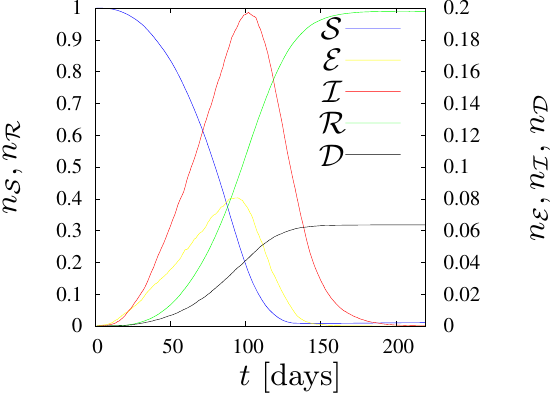}
\end{subfigure}
\begin{subfigure}[b]{.329\textwidth}
\caption{\label{fig:pE0.03_pI0.02_rI1.0_rE2.5}$r_\mathcal{E}=2.5$}
\includegraphics[width=.94\columnwidth]{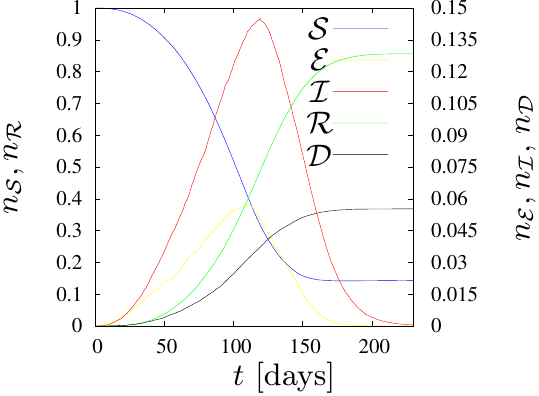}
\end{subfigure}
\begin{subfigure}[b]{.329\textwidth}
\caption{\label{fig:pE0.03_pI0.02_rE3.0_rI2.5}$r_\mathcal{I}=2.5$}
\includegraphics[width=.94\columnwidth]{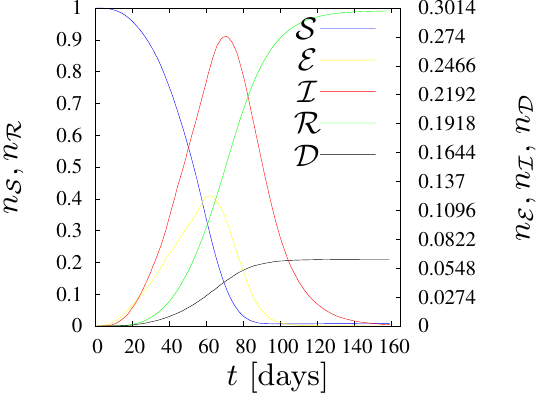}
\end{subfigure}
\begin{subfigure}[b]{.329\textwidth}
\caption{\label{fig:pE0.03_pI0.02_rErI3.0}$r_\mathcal{I}=r_\mathcal{E}=3$}
\includegraphics[width=.94\columnwidth]{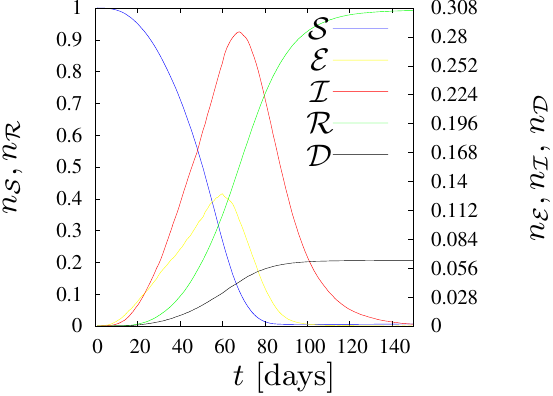}
\end{subfigure}
\begin{subfigure}[b]{.329\textwidth}
\caption{\label{fig:pE0.03_pI0.02_rI1.0_rE3.0}$r_\mathcal{E}=3$}
\includegraphics[width=.94\columnwidth]{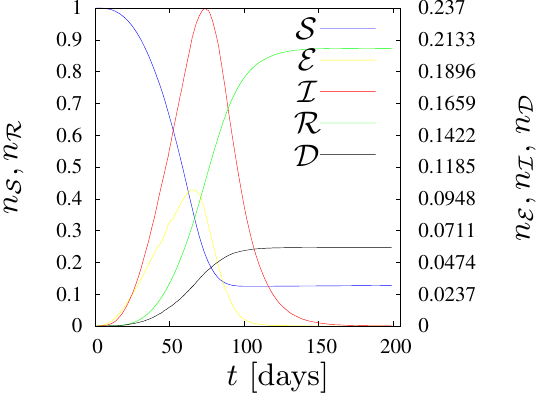}
\end{subfigure}
\begin{subfigure}[b]{.329\textwidth}
\caption{\label{fig:pE0.03_pI0.02_rE3.0_rI3.0}$r_\mathcal{I}=3$}
\includegraphics[width=.94\columnwidth]{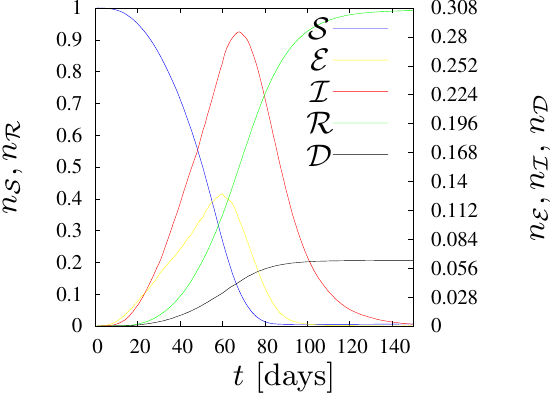}
\end{subfigure}
\caption{\label{fig:pE0.03_pI0.02}Dynamics of states fractions for various values of neighbourhood radius
  (left) $r_\mathcal{E}=r_\mathcal{I}$,
(middle) $r_\mathcal{E}\ge r_\mathcal{I}=1$,
 (right) $3=r_\mathcal{E}\ge r_\mathcal{I}$.
$p_\mathcal{E} = 0.03$, $p_\mathcal{I} = 0.02$, $R=10$.}
\end{figure*}

In the left column of \Cref{fig:pE0.03_pI0.02} the dynamics of states fractions $n$ 
for various values of the radius of the neighborhood $r_\mathcal{E}=r_\mathcal{I}$ are presented. 
We assume $p_\mathcal{E}=0.03$, $p_\mathcal{I} = 0.02$. 
As we still keep $r_\mathcal{I} = r_\mathcal{E}$ setting $p_{\mathcal E}>p_{\mathcal I}$ simulates the fact that exposed agents (who are not aware of their infection) are more dangerous to those around them than those who know that they are sick, and therefore avoid contact if they show severe symptoms of the disease.

In \Cref{fig:pE0.03_pI0.02_rErI1.0} the results for the smallest possible neighborhood radius $r>0$ are presented.
As in the previous cases with $r=1$, the epidemic stops quite quickly. The largest number of agents who are ill ($n_{\mathcal I}L^2$) at the same time is two, the longest simulation time is 120 days, the average number of deaths is well below one, and totally only a few agents are infected (the curve representing agents in the state of $\mathcal R$ is almost not visible). 

The \Cref{fig:pE0.03_pI0.02_rErI1.5} shows that contact with eight agents in the neighborhood (for assumed values of infection rates $p_{\mathcal E}$ and $p_{\mathcal I}$) is enough for the pandemic to affect society to a very large extent and last for a long time. The longest simulation took about 1000 days. During the epidemic, almost half of the population was infected, approximately 350 agents died, while at epidemic peak there were just over 180 sick agents in a single day.

Increasing the range of interaction to the next coordination zone (see \Cref{fig:pE0.03_pI0.02_rErI2.0}) caused a significant reduction in the duration of the pandemic, as well as the greater havoc it caused among agents. In approximately 350 days, roughly 75\% of the population was infected, approximately 500 died, and up to 800 were sick on the day when this number was highest.

Increasing the radius to 2.5, as shown in \Cref{fig:pE0.03_pI0.02_rErI2.5}, further shortened the epidemic time, this time to just two hundred days. Almost all agents were infected, slightly more than 600 agents died, and the maximum number of patients on one day was 2000, which is as much as 20\% of the population. Every fifth person was infected by the disease on that day, and if you add about 800 agents who were in exposed state $\mathcal E$, we get a situation when more than one-quarter of the population is under the influence of this disease at the same time.

Further increasing the range of interactions (to the fifth coordination zone and 24 neighbors in it, shown in \Cref{fig:pE0.03_pI0.02_rErI3.0}) gives very similar results to the previous one, except that the virus spreads even faster, in less than 150 days. The number of deaths is similar (a little below 600), almost all agents had contact with the disease at some stage of the pandemic, and the maximum number of sick agents in one day was 2800. If you add over 1100 agents in the $\mathcal E$ state on the same day, the effect is that the disease affected nearly 40\% of agents on the same day.

\subsection{\label{sec:rE_gt_rI}$r_\mathcal{E}\ge r_\mathcal{I}$}

In the middle column of \Cref{fig:pE0.03_pI0.02} the dynamics of state fractions $n_\mathcal{S}$, $n_\mathcal{E}$, $n_\mathcal{I}$, $n_\mathcal{R}$ 
for various values of the neighborhood radius $r_\mathcal{E}$ are presented.
We assume $p_\mathcal{E}=0.03$ and $p_\mathcal{I}=0.02$. 
For this set of plots, $r_{\mathcal I}$ has been predefined as 1 and only $r_{\mathcal E}$ changes. 
Translating this into a description of the real world, sick agents who have disease symptoms are aware of their illness (infected) and limit their contact with the environment to a minimum, while oblivious agents (exposed) do not realize that they can transmit the disease.
In each of the tested parameter sets, agents in $\mathcal I$ state can infect only four of their closest neighbors.

The case illustrated in \Cref{fig:pE0.03_pI0.02_rI1.0_rE1.0} was analyzed in the \Cref{sec:rE_eq_rI} in \Cref{fig:pE0.03_pI0.02_rErI1.0}.

The plot in  \Cref{fig:pE0.03_pI0.02_rI1.0_rE1.5} shows the situation where agents in the state $\mathcal E$ can infect up to eight agents in their neighborhood. We see irregularities in the shapes of the curves showing the fraction of agents in the state $\mathcal I$ and $\mathcal E$, the duration of the epidemic is relatively long, over 400 days. However, the values shown in the graphs show that the pandemic was not dangerous for the entire population.
Death from the disease was recorded on average in less than eight agents, the maximum number of sick agents in one day is just over six, while the total number of infected agents in the entire epidemic was so low that the deviation of the curve representing the number of agents in the state $\mathcal S$ from the top of the graph is almost imperceptible.

In \Cref{fig:pE0.03_pI0.02_rI1.0_rE2.0} the plots for further extension (to the third coordination zone, with $z=12$ neighbors) of the radius of the neighborhood of exposed agents (in the state $\mathcal E$) are presented. 
The curves are much smoother than those presented in \Cref{fig:pE0.03_pI0.02_rI1.0_rE1.5}, however, we can still see some irregularities in the curves describing agents in the $\mathcal E$ and $\mathcal I$ states. The duration of the longest simulation was approximately 700 days, the number of deaths was less than 330, and at the peak of the pandemic, approximately 220 agents were simultaneously ill. The curves presenting the fraction of agents in the states of $\mathcal S$ and $\mathcal R$ almost perfectly line up on the right side of the chart, meaning that half of the population contracted the disease while the other half remained healthy throughout the epidemic.

The case where agents in the state $\mathcal E$ can infect agents within a radius of 2.5 is shown in \Cref{fig:pE0.03_pI0.02_rI1.0_rE2.5}. 
The epidemic was definitely more dynamic than in the previously studied case (\Cref{fig:pE0.03_pI0.02_rI1.0_rE2.0}), lasting only a little over two hundred days and the maximum number of sick agents in one day $n_IL^2$ reached almost 1500 agents. The sum of those who died from the coronavirus is less than 600, and during the pandemic, approximately 85\% of the population became recovered (and earlier exposed and/or infected).

Finally, for the radius $r_{\mathcal E}$ increased to three (see \Cref{fig:pE0.03_pI0.02_rI1.0_rE3.0}) we do not observe too many changes compared to the previous case (\Cref{fig:pE0.03_pI0.02_rI1.0_rE2.5}). The epidemic is even shorter, it lasts a little more than 150 days, the number of deaths is very similar (it can be estimated at 600), slightly more than 85\% of agents in the population were infected, while at the hardest moment of the pandemic, there were at the same time about 2,370 sick agents (in $\mathcal I$ state). The results are therefore almost identical to those of $r_{\mathcal E}=2.5$, except that disease propagation is faster.

The right column of \Cref{fig:pE0.03_pI0.02} shows cases with predetermined $r_{\mathcal E}=3$ and various $r_{\mathcal I}$.
We keep $p_\mathcal{E} = 0.03$ and $p_\mathcal{I} = 0.02$

All graphs are very similar to each other, as a large range of infections of agents in the state $\mathcal E$ makes influence of $r_{\mathcal I}$ on epidemic evolution only marginal. It can be easily observed by comparing the scales presented in all subfigures. The assumed range of infection $r_{\mathcal E}=3$ is large enough  (at least for assumed values of transmission rates $p_{\mathcal E}$ and $p_{\mathcal I}$) to prevent any observable influence of $r_{\mathcal I}$ on epidemic trajectories.

\section{\label{sec:discussion}Discussion}

Let us start the discussion with comparing the most left (\Cref{fig:pE0.03_pI0.02_rErI1.0,fig:pE0.03_pI0.02_rErI1.5,fig:pE0.03_pI0.02_rErI2.0,fig:pE0.03_pI0.02_rErI2.5,fig:pE0.03_pI0.02_rErI3.0}) and the middle (\Cref{fig:pE0.03_pI0.02_rI1.0_rE1.0,fig:pE0.03_pI0.02_rI1.0_rE1.5,fig:pE0.03_pI0.02_rI1.0_rE2.0,fig:pE0.03_pI0.02_rI1.0_rE2.5,fig:pE0.03_pI0.02_rI1.0_rE3.0}) columns of \Cref{fig:pE0.03_pI0.02}. 
The comparison reveals that quarantining or limiting the connectivity of agents in the $\mathcal I$ state (both, infected and informed) may bring good or even very good results in preventing disease propagation, depending on the arrangement of the other parameters. For $r_{\mathcal E}=1.5$, this completely brought the epidemic to a halt, which would otherwise affect more than half the population. With $r_{\mathcal E}=2$, it makes it possible to reduce the share of infected agents in the population from 75\% to 50\%. The effects on $r_{\mathcal E}=2.5$ and $r_{\mathcal E}=3$ are similar --- instead of the total population less than 90\% of population became infected. When $r_{\mathcal E}=1.5$ pandemic duration was significantly shortened, it is because the virus was unable to survive, and the disease was extinct. In other cases, the duration of the epidemic increased, the restrictions introduced for agents in the $\mathcal I$ state did not completely extinguish the disease, but allowed to slow it down and mitigate its effects.

For low values of the probability of infection ($p_{\mathcal E}=p_{\mathcal I}=0.005$), that is, in a situation where the transmission of the virus is not too high, even a slight limitation of the contact among agents allows for a complete inhibition of the disease and protection of the society against its negative effects (see \Cref{fig:pE0.005_pI0.005_rErI}).
We note that manipulating in $p_{\mathcal E}$ and/or $p_{\mathcal I}$ parameters may reflect changes in disease transition rates with their low values corresponding to wild variant of SARS-CoV-2 virus while higher values to the fiercer (including delta and specially omicron) variants of SARS-CoV-2 virus.
For instance, for a fixed radius of interaction ($r_{\mathcal I} = r_{\mathcal E} = 3$) for low values of $p_{\mathcal E} = p_{\mathcal I} = 0.005$ nearly 75\% of population reached the $\mathcal R$ state (\Cref{fig:pE0.005_pI0.005_rErI3.0}) while increasing infection rates to $p_{\mathcal E} = 0.03$ and $p_{\mathcal I} = 0.02$ caused the entire population to fall ill (\Cref{fig:pE0.03_pI0.02_rErI3.0}).
The summed values of fractions $n_{\mathcal E}$ of exposed and $n_{\mathcal I}$ infected agents at the peaks of disease are 6.6\% (for $p_{\mathcal E} = p_{\mathcal I} = 0.005$, \Cref{fig:pE0.005_pI0.005_rErI3.0}) and 39\% (for $p_{\mathcal E} = 0.03$ and $p_{\mathcal I} = 0.02$, \Cref{fig:pE0.03_pI0.02_rErI3.0}) of the population.

\begin{figure}[htbp]
\centering
\captionsetup{justification=centering}
\includegraphics[width=.99\columnwidth]{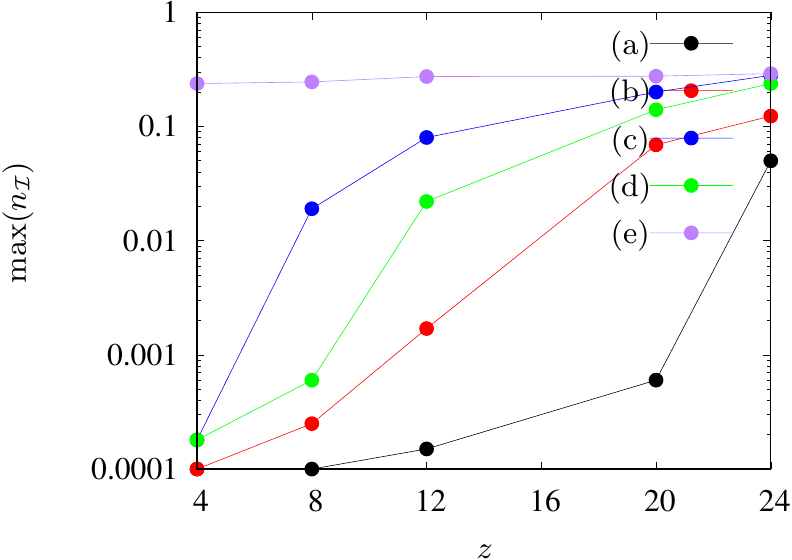}
\caption{\label{fig:maxIvsr}Maximal fraction $n_{\mathcal I}$ of agents in state $\mathcal I$ as dependent on the number of agents' neighbours $z$ in the neighbourhood.
  (a) $p_{\mathcal E}=p_{\mathcal I}=0.005$, $z_E=z_{\mathcal I}=z$,
  (b) $p_{\mathcal E}=p_{\mathcal I}=0.01$,  $z_E=z_{\mathcal I}=z$,
  (c) $p_{\mathcal E}=0.03$, $p_{\mathcal I}=0.02$, $z_E=z_{\mathcal I}=z$,
  (d) $p_{\mathcal E}=0.03$, $p_{\mathcal I}=0.02$, $z_E=z$, $z_{\mathcal I}=4$,
  (e) $p_{\mathcal E}=0.03$, $p_{\mathcal I}=0.02$, $z_E=24$, $z_{\mathcal I}=z$.}
\end{figure}

The results presented in \Cref{fig:pE0.005_pI0.005_rErI,fig:pE0.03_pI0.02} are summarized in \Cref{fig:maxIvsr}, where the maximum fraction of agents in state $\mathcal I$ vs. the neighbours number $z$ for various sets of parameters is presented.
We observe a gradual increase in the maximum number of infected agents (up to 30\%) as the number of agents in the neighborhood increases. An exception to this rule is observed only when $r_\mathcal{E} = 3$, when the maximum level of infections is constant and it does not change with the increase of the range of the interaction of agents in state $\mathcal{I}$.

\section{\label{sec:conclusions}Conclusions}

In this article we present a stochastic synchronous cellular automaton defined on a square lattice.
The automaton rules are based on the SEIR model with probabilistic parameters collected from real human mortality data and SARS-CoV-2 disease characteristics.
Automaton rules are presented in \Cref{alg:SEIR}.
With computer simulations, we show the influence of the radius of the neighborhood on the number of infected and deceased agents in the artificial population.

The study presented in this paper is based on static automaton. Thus our approach is equivalent to disease propagation described in terms of conduction-like processes (i.e. the position of each cell is fixed and can infect a neighbor, at distance $r_\mathcal{E}$ or $r_\mathcal{I}$). The latter is a natural way to model with the cellular automata technique. However, we note that also convection-like processes (i.e. the population can flow within the system) may play a crucial role at both large \cite{PhysRevX.1.011001} and local scale \cite{PhysRevResearch.2.043379,Mello_2020}.

Further enrichment of the model can lead to the introduction of additional components, including the compartment $\mathcal{V}$ that describes the vaccinated agents. This model improvement allows us to study scenarios with limited and unlimited vaccine supply \cite{Gabrick_2022} or the existence and stability of steady states \cite{Sun_2021,Nabti_2021}.
Vaccinations seem to be particularly effective when the vaccination campaign starts early and with a large number of vaccinated individuals \cite{Gabrick_2022}.
Moreover, the realistic models should take into account people's attitudes to vaccination programs \cite{Bier_2015,Lisowski_2019}.

In conclusion, increasing the radius of the neighborhood (and the number of agents interacting locally) favors the spread of the epidemic.
However, for a wide range of interactions of exposed agents, even isolation of infected agents cannot prevent successful disease propagation.
This supports aggressive testing against disease as one of the useful strategies to prevent large peaks of infection in the spread of SARS-CoV-2-like disease. 
The latter can have devastating consequences for the health care system, in particular for the availability of hospital beds for SARS-CoV-2 and other diseases.

\begin{acknowledgements}
Authors are grateful to Zdzis{\l}aw Burda for critical reading of the manuscript and fruitful discussion.
\end{acknowledgements}

%

\end{document}